\documentclass{amsproc}
\usepackage{graphicx}
\usepackage{amsmath}
\usepackage{amssymb}
\usepackage{subfigure}
\usepackage[all]{xy}  % for long exact sequences
\usepackage{color}
\usepackage{mathtools}
\usepackage{cite}

\usepackage{hyperref}
\hypersetup{pdftitle={Target Space Duality of Heterotic GUTs}, 
      pdfauthor={Thorsten Rahn}, 
      pdfsubject={Algebraic Geometry, Mathematical Methods in Physics},
      pdfstartview={FitH}, pdfpagelayout={TwoColumnRight},
      bookmarksopen, bookmarksnumbered, bookmarksopenlevel=2}
\definecolor{darkblue}{rgb}{0,0,.5}
\definecolor{lightblue}{rgb}{0.8,0.85,1}
\definecolor{mygray}{gray}{.75}

\newcommand{\cohomCalgKoszul}{{\text{\fontfamily{put}\bfseries\footnotesize\selectfont cohomCalg Koszul} extension}}

\newcommand{\eq}[1]{\begin{equation}
                     \begin{split} #1 \end{split}
                     \end{equation}}
\newcommand{\beq}{\begin{equation}}  \newcommand{\eeq}{\end{equation}}
\newcommand{\bal}{\begin{aligned}}   \newcommand{\eal}{\end{aligned}}

\def\IP{\mathbb{P}}

\def\cM{\mathcal{M}}
\def\cO{\mathcal{O}}

\def\cV{\mathcal{V}}

\def\clap#1{\hbox to 0pt{\hss#1\hss}}
\def\mllap{\mathpalette\mathllapinternal}

\def\mathllapinternal#1#2{%
\llap{$\mathsurround=0pt#1{#2}$}}

\def\uspc{${}^\big.$}   % ugly "spaceholders" for table formatting

\begin{document}
% 
% \baselineskip=14pt
% %\parskip 3pt plus 1pt 
% 
 \vspace*{-4cm}
 \begin{flushright}    % Publication numbers
   {\small  MPP-2011-131}
 \end{flushright}
 \vspace*{4cm}
% 
% \vspace{2cm}
% \begin{center}        % Main title
%   {\LARGE
%   Target Space Dualities of Heterotic Grand Unified Theories%$E_6,~SO(10),~SU(5)$ GUTs 
% %Landscape Study of  Target Space Duality of  \\[0.4cm]    $(0,2)$ Heterotic String Models    
%   }
% \end{center}
% 
% \vspace{0.75cm}
% \begin{center}        % Authors
%   Thorsten Rahn
% \end{center}
% 
% \vspace{0.15cm}
% \begin{center}        % Institutes 
%   \emph{Max-Planck-Institut f\"ur Physik, F\"ohringer Ring 6, \\ 
%                80805 M\"unchen, Germany} \\[5mm]
% \end{center} 
% 
% \vspace{2cm}

%%%%%%%%%%%%%%%%%%%%%%%%%%%%%%%%%%%%%%%%%%%%%%%
%%%%%%%%%%%%%%%%%%%%%%%%%%%%%%%%%%%%%%%%%%%%%%%
%%%%%%%%%%%%%%%%%%%%%%%%%%%%%%%%%%%%%%%%%%%%%%%
%%%%%%%%%%%%%%%%%%%%%%%%%%%%%%%%%%%%%%%%%%%%%%%
%%%%%%%%%%%%%%%%%%%%%%%%%%%%%%%%%%%%%%%%%%%%%%%
%%%%%%%%%%%%%%%%%%%%%%%%%%%%%%%%%%%%%%%%%%%%%%%
%%%%%%%%%%%%%%%%%%%%%%%%%%%%%%%%%%%%%%%%%%%%%%%
%%%%%%%%%%%%%%%%%%%%%%%%%%%%%%%%%%%%%%%%%%%%%%%
% \title[short text for running head]{full title}
\title{Target Space Dualities of Heterotic Grand Unified Theories}
%
%    Only \author and \address are required; other information is
%    optional.  Remove any unused author tags.
%
%    author one information
% \author[short version for running head]{name for top of paper}
\author{Thorsten Rahn}
\address{Max-Planck-Institut f\"ur Physik, F\"ohringer Ring 6, 80805 M\"unchen, Germany}
%\curraddr{}
\email{rahn@mpp.mpg.de}
%\thanks{}
%
%\subjclass[2000]{Primary }
%    The 2010 edition of the Mathematics Subject Classification is
%    now available.  If you are citing a classification from the
%    new scheme, use the following input coding instead.
%\subjclass[2010]{Primary }
%
%\date{}
%
\begin{abstract}
In this article we summarize and extend the ideas and investigations on so called target space dualities of heterotic models with $(0,2)$ worldsheet supersymmetry as they were partly presented on the String-Math 2011 conference. After the generic description of the duality, we give some novel examples involving vector bundles that are not deformations of the tangent bundle but more generic ones corresponding to $SO(10)$ and $SU(5)$ gauge theories in four dimensions. We show explicitly that the necessary conditions for a duality also hold for compactifications of this kind. Finally we will present the results of the large landscape scan of $E_6$ models.
\end{abstract}
\maketitle
\section{Introduction}
Heterotic string theory provides a way to build grand unified theories very naturally. A priori it arises with an $E_8\times E_8$ gauge group which can then be broken down to some GUT group in four dimensions. In such a model, on the one hand, we have bosonic degrees of freedom that are valued in the Calabi-Yau manifold, denoted by $\cM$ throughout this paper, and on the other hand their fermionic superpartners that are coupled to the pullback of the tangent bundle. Furthermore we have some left-moving fermions that do not necessarily couple to the pullback of the tangent bundle but more generically to the pullback of some vector bundle $\cV$ of rank $n=3,4$ or $5$. The only conditions on this vector bundle is that it is holomorphic, stable and that the first an second Chern classes equal those of the tangent bundle. The structure group of $\cV$ is then $SU(3),~SU(4)$ or $SU(5)$ for a rank $3,~4$ or $5$ bundle respectively and breaks one of the $E_8$ factors down to the commutant of $SU(n)$ in $E_8$ which gives an effective four-dimensional theory with gauge group $E_6,~SO(10)$ or $SU(5)$ respectively whereas the other $E_8$ factor can be hidden. In the geometric picture the massless matter spectrum can be obtained by computing certain vector bundle valued cohomology classes. In case of an $E_6,~SO(10)$ or $SU(5)$ gauge group the actual cohomologies that need to be computed  can be found in table \ref{table_GUT group representations via cohomology}. A review on the computational tools for such a kind of calculation can be found e.g.~in  \cite{Blumenhagen:2011xn} and references therein.
\begin{table}[ht]
  \small\newcommand{\Xoplus}{\mllap{\oplus\,}}
  \begin{tabular}{l|cccccc}
    \# zero modes & &&&&&\\
    in reps of $H\times G$&  1 &   $h^1_{\mathcal M}({\mathcal V})$ & $h^1_{\mathcal M}({\mathcal V}^\ast)$ &  $h^1_{\mathcal M}(\Lambda^2 {\mathcal V})$ &  $h^1_{\mathcal M}(\Lambda^2 {\mathcal V}^\ast)$ & $h^1_{\mathcal M}({\mathcal V}\otimes {\mathcal V}^\ast)$  \\
    \hline\hline
    \qquad\;\;\,\parbox{1cm}{$E_8$\uspc \\ ${}\,\downarrow$} & \multicolumn{6}{|c}{\parbox{1cm}{248\uspc \\ ${}\;\downarrow$}}  \\ 
    $SU(3) \times E_6$     &  $(1,78)$ & $\Xoplus(3,27)$ & $\Xoplus(\overline{3},\overline{27})$  & & & $\Xoplus(8,1)$\uspc \\
    $SU(4) \times SO(10)$  & $(1,45)$ & $\Xoplus(4,16)$ & $\Xoplus(\overline{4},\overline{16})$ & $\Xoplus(6,10)$ & & $\Xoplus(15,1)$\\
    $SU(5) \times SU(5)$   &  $(1,24)$ & $\Xoplus(5,\overline{10})$ & $\Xoplus(\overline{5}, 10)$ & $\Xoplus(10,5)$ & $\Xoplus(\overline{10},\overline{5})$ & $\Xoplus (24,1)$\\
  \end{tabular}
  \caption{\small Matter zero modes in representations of the GUT group }
  \label{table_GUT group representations via cohomology}
\end{table}

One nice way to see how holomorphic vector bundles arise in string-theory  is using the $(0,2)$ gauged linear sigma model (GLSM) \cite{Witten:1993yc}. Here one starts with a two dimensional gauge  theory containing certain chiral as well as Fermi superfields. The most general superpotential of such a theory can be written as
\begin{equation}
\label{ftermsuperpot}
 S=\int d^2 z d\theta\,  \left[ \sum_j \Gamma^j\,  G_j(X_i) + 
               \sum_{l,a} P_l\,  \Lambda^a\,  F_a{}^l(X_i)  \right]\,,
\end{equation}
where the $X_i,~P_l$ and $\Gamma^j,~\Lambda^a$ correspond to chiral and Fermi superfields. They are all charged under a certain number of $U(1)$ gauge groups in such a way that the superpotential is gauge invariant. If we denote the bosonic components of the chiral superfields $X_i$ and $P_l$ by $x_i$ and $p_l$ we will find a bosonic potential for them consisting of an F-term potential
\begin{equation}
  V_F= \sum_j \Bigl\vert G_j(x_i)\Bigr\vert^2  + 
               \sum_a \Bigl\vert \sum_{l} p_l\,   F_a{}^l(x_i)  \Bigr|^2 
\end{equation}
as well as a D-term scalar potential
\begin{equation}
  V_D= \sum_{\alpha=1}^r   \biggl(  \sum_{i=1}^ d  Q_i^{(\alpha)} \vert x_i \vert^2 - \sum_{l=1}^\gamma
     M_l^{(\alpha)} \vert p_l \vert^2 -\xi^{(\alpha)} \biggr)^2 \,,
\end{equation}
where the $\xi^{(\alpha)}\in \mathbb R$ denote the Fayet-Iliopoulos (FI) parameters for each $U(1)$ gauge group. One can now go ahead and analyze the vacuum structure of such a theory and certainly this structure will crucially depend on the actual value of the FI parameters. Putting more weight on the mathematical point of view, one calls them also K\"ahler parameters. Different choices of these parameters can be interpreted as different triangulations of some polytope and each triangulation describes one kind of vacuum or phase of the underlying GLSM \cite{Aspinwall:1993nu}. If one chooses them in such a way that the triangulation is maximal, i.e.~it has more or the same number of maximal dimensional cones than every other possible triangulation then our vacuum may correspond to a holomorphic vector bundle over a Calabi-Yau manifold. In the low energy effective action this will be the corresponding target space of the non-linear sigma model.

Here the idea of target space dualities comes into play. It basically works with the phases of the GLSM that do not have a completely geometric interpretation and uses some freedom in them to change the GLSM without spoiling the phase itself. This basically means that the moduli spaces of two GLSMs can be connected and that there exists a locus inside a specific phase where they coincide. After the redefinition of the data, we can just pretend that we came from the new GLSM and go back to the geometric phase there. The question is now whether we actually undergo a transition of one geometry to a different one by doing this or if we just walk around in one and the same moduli space and are dealing with a target space duality between those two geometries. Distler and Kachru first pointed out that such a thing might actually exist \cite{DistlerKachruDuality} and further work \cite{DistlerGreeneFeatures, RalphTargetSpace1,RalphTargetSpace2} supported this idea with specific examples by explicitly comparing the dimensions of the moduli spaces as well as the spectrum of the corresponding models and found agreement:
\begin{eqnarray}
     h^{1,1}({\mathcal M}) + h^{2,1}({\mathcal M}) +  h^{1}_{\mathcal M}({\rm
       End}({\mathcal V})) &=&
     h^{1,1}(\widetilde{\mathcal M}) + h^{2,1}(\widetilde{\mathcal M}) +  h^{1}_{\widetilde{\mathcal M}}({\rm
       End}(\widetilde{\mathcal V})) \;,\nonumber\\
h^{i}_{\mathcal M}(\wedge^k\, {\mathcal V})  &=& h^{i}_{\widetilde{\mathcal M}}(\wedge^k\,
\widetilde{\mathcal V}),\qquad  {\rm for}\  \ i = 0,\ldots, 3 \nonumber\,.
\end{eqnarray}
%%%%%%%%%%%%%%%%%%%%%%%%%%%%%%%%%%%%%%%%%%%%%%%
%%%%%%%%%%%%%%%%%%%%%%%%%%%%%%%%%%%%%%%%%%%%%%%
\section{The Duality}
%%%%%%%%%%%%%%%%%%%%%%%%%%%%%%%%%%%%%%%%%%%%%%%
%%%%%%%%%%%%%%%%%%%%%%%%%%%%%%%%%%%%%%%%%%%%%%%
In this section we will review the general procedure that can be employed to produce dual models from a given one. In order to do that we first explain how this can be done schematically and then turn to three explicit and novel examples. Here we will capture all possible structure groups for the bundle, i.e.~ we consider an $SU(3)$, $SU(4)$ and at last an $SU(5)$ bundle. That the duality works for such models was mentioned in \cite{LandscapeStudy} but not shown in an explicit example. We want to use the opportunity to catch up on that. 
\subsection{General procedure}\label{subsec_construction of dual models} 
In the setting of the GLSM we mentioned above we can specify the vector bundle as the cohomology of a complex containing direct sums of line bundles
\begin{equation}\label{eq_general monad}
 0\rightarrow \cO_{\cM}^{\oplus r_{\cV}}\;  {\buildrel \otimes E_i{}^a \over
   \longrightarrow}\;  \bigoplus_{a=1}^{\delta} \cO_\cM(N_a) \; {\buildrel \otimes F_a{}^l
   \over \longrightarrow}\;  \bigoplus_{l=1}^{\gamma} \cO_\cM(M_l)\rightarrow 0\,.
\end{equation}
This means that the vector bundle is given by
\beq
\cV = \frac{\text{ker}(F_a{}^l)}{\text{im}(E_i{}^a)}\,.
\eeq
The chiral and Fermi fields of the corresponding GLSM are charged under the various $C^\ast$ actions of the toric variety that contains $\cM$ as complete intersection of hypersurfaces. In the following we will denote the defining data of the configuration by two tables
\begin{equation*}
\begin{aligned}
&\begin{array}{|c||c|}
\hline
x_i & \Gamma^j \\
\noalign{\hrule height 1pt}
\begin{array}{ccc}
 Q_1 & \cdots & Q_N
\end{array}
&
\begin{array}{ccc}
-S_1 & \cdots & -S_c
 \end{array}\\
\hline
\end{array}
&\hspace{-0.25cm}
\begin{array}{|c||c|}
\hline
\Lambda^a & p_l \\
\noalign{\hrule height 1pt}
\hline
\begin{array}{ccc}
 N_1 & \cdots & N_\delta
\end{array}
&
\begin{array}{ccc}
  -M_1 & \cdots & -M_\gamma
\end{array}\\
\hline
\end{array}
\end{aligned}\,.
\end{equation*}
In this scenario we can turn the requirement of matching first and second Chern classes of $\cV$ and $T_{\cM}$ into a combinatorial relation of the charges of the fields:
\begin{equation}\label{eq_anomcancel}
\begin{aligned}
  \sum_{a=1}^\delta  N_a^{(\alpha)} = \sum_{l=1}^\gamma  M_l^{(\alpha)},\qquad\quad
   &\sum_{i=1}^d  Q_i^{(\alpha)} = \sum_{j=1}^c  S_j^{(\alpha)}\,, \\ %\label{eq_anomaly cancellation 3}
  \sum_{l=1}^\gamma  M_l^{(\alpha)} M_l^{(\beta)} -   \sum_{a=1}^\delta  N_a^{(\alpha)} N_a^{(\beta)} 
  =&  \sum_{j=1}^c  S_j^{(\alpha)}  S_j^{(\beta)}  - \sum_{i=1}^d Q_i^{(\alpha)} Q_i^{(\beta)} \;.
\end{aligned}
\end{equation}
Here the Greek index in parenthesis labels the corresponding $C^\ast$ action of the toric variety and the equations \eqref{eq_anomcancel} have to be satisfied for all $\alpha$ and $\beta$.
Now after fixing the notations we can go on to describe the explicit procedure which consists of different steps:\\
\paragraph{\bf The procedure:}
\begin{enumerate}
 \item Construct the GLSM phases of a smooth $(0,2)$ model
    $({\mathcal M},{\mathcal V})$.
 \item Go to a phase where one of the $p_{l}$, say $p_{1}$, is not allowed to vanish and hence obtains a vev $\left< p_{1} \right>$.
 \item Perform a rescaling of $k$ Fermi superfields by the constant 
vev $\left< p_{1} \right>$ and exchange the role 
of some $\Lambda^a$ and $\Gamma^j$
       $$\tilde\Lambda^{a_i} :={\frac{\Gamma^{j_i}}{\left< p_{1} \right>}}, \quad
\tilde\Gamma^{j_i}  := \left< p_{1} \right>\Lambda^{a_i}\,,\quad ~\forall i=1,...,k\,,$$
      with  $\sum_{i} || G_{j_i} || = \sum_{i} || F_{a_i}{}^{1} ||$ for anomaly cancellation.
 \item Move to a region in the bundle moduli space where the $\Lambda^{a_i}$
   only appear in terms with $P_{1}$ for all $i$. This means that  
       we choose the coefficients in the bundle defining polynomials 
$F_{a}{}^l$ such that $$F_{a_i}{}^l=0\, , 
  \qquad\forall~l\neq 1,\ ~i=1,...,k\,.$$
 \item Leave the non-geometric phase and define the
   Fermi superfields of the new GLSM 
such that each term in the superpotential is $U(1)^r$ gauge invariant. This means
 $$||\tilde\Lambda^{a_i}|| = ||\Gamma^{j_i}||-||P_{1}|| \quad\text{and}\quad
||\tilde\Gamma^{j_i}|| = ||\Lambda^{a_i}|| + ||P_{1}||\, . $$
 \item Returning to a generic point in moduli space defines a new dual 
$(0,2)$ GLSM which in a geometric phase corresponds to a different
 Calabi-Yau/vector bundle configuration $(\widetilde {\mathcal M},\widetilde {\mathcal V})$.
\end{enumerate}
\subsection{Examples of models with structure group SU(n)}
We now want to be a bit more explicit and show some specific examples of the duality. We will show one example for each type of structure group $SU(3)$, $SU(4)$ and $SU(5)$. In contrast to earlier work \cite{LandscapeStudy} where we mostly focused on models that were deformations of the tangent bundle and hence given by the cohomology of the Euler sequence, here we want to give different examples that do arise from an exact monad
\begin{equation}\label{eq_exact monad}
 0\rightarrow {\cV}\;  {\buildrel f \over
   \longrightarrow}\;  \bigoplus_{a=1}^{\delta} \cO_\cM(N_a) \; {\buildrel \otimes F_a{}^l
   \over \longrightarrow}\;  \bigoplus_{l=1}^{\gamma} \cO_\cM(M_l)\rightarrow 0\,,
\end{equation}
and hence are given by the kernel of the map $F_a{}^l$
\beq
 \cV = \text{Ker}\left( F_a{}^l \right)\,.
\eeq
The way to generate the dual models of such a monad remains the same.\\
%`
\paragraph{\bf Example for an \boldmath{ $SU(3)$}-model:}
We start with an $SU(3)$ example which consists of a holomorphic vector bundle over a codimension two complete intersection Calabi-Yau space. In \cite{LandscapeStudy} we investigated mostly $SU(3)$ models which are given by a deformation of the tangent bundle. As described there the base space undergoes usually a conifold transition. In this example here we are not dealing with a $(0,2)$ model which is a deformation of the tangent bundle but a completely independent monad. Furthermore we will see that the base will not transform via a conifold transition. Rather in the beginning the ambient variety will remain untouched and only a different set of hypersurfaces will be chosen, also resulting in a topology change of the base. Finally through the exchange of those specific hypersurfaces we will see that in fact the ambient space topology will be changed after all. The model data is given by
\begin{equation}
\begin{aligned}
&\begin{array}{|c||c|}
\hline
x_i & \Gamma^j \\
\noalign{\hrule height 1pt}
\begin{array}{ccccccc}
 0 & 0 & 0 & 1 & 1 & 1 & 1 \\
 1 & 1 & 1 & 2 & 2 & 2 & 0
\end{array}
&
\begin{array}{cc}
 -2 & -2  \\
 -4 & -5 
 \end{array}\\
\hline
\end{array}
\quad
&\begin{array}{|c||c|}
\hline
\Lambda^a & p_l \\
\noalign{\hrule height 1pt}
\hline
\begin{array}{cccccccc}
 1 & 0 & 0 & 2  \\
 0 & 1 & 1 & 6 
\end{array}
&
\begin{array}{c}
 -3 \\
 -8  
\end{array}\\
\hline
\end{array}\,.
\end{aligned}
\end{equation}
\normalsize
\noindent
As was explained in  \cite{Blumenhagen:2010ed} to compute the number of chiral matter zero modes and the massless singlets, we need nothing but line bundle cohomologies. Furthermore the efficient algorithm suggested in \cite{Blumenhagen:2010pv} and proven in \cite{Rahn:2010fm} and \cite{2010arXiv1006.0780J} allows one to calculate such cohomologies quite fast. Employing our implementation \cohomCalgKoszul \cite{cohomCalg:Implementation} for this matter we find
\eq{
\label{eq_match chiral spectrum example 1}
h_{\mathcal M}^\bullet(\mathcal V) &= ( 0, 120 , 0,0)\,,\\
 h_{\mathcal M}^{1,1}+h_{\mathcal M}^{2,1}+h_{\mathcal M}^1 (\text{End}({\mathcal V})) &=  2 + 68 + 322 = 392\,.
}
In order to see our freedom of consistently exchanging hypersurface equations with bundle maps in the monad as described in the last section we explicitly write down the multi-degrees of the corresponding generic homogeneous functions. Using that
\beq
 ||F_{a}{}^l|| = -||p_l|| - ||\Lambda^a||\,,
\eeq
for the oly choice $l=1$ they read
\begin{eqnarray}
 ||G_1||=\left(\!\!\begin{array}{c}2 \\ 4\end{array}\!\!\right)\,,
 ||G_2||=\left(\!\!\begin{array}{c}2 \\ 5\end{array}\!\!\right)\,,\\
 ||F_1{}^1||=\left(\!\!\begin{array}{c}2 \\ 8\end{array}\!\!\right)\,,
 ||F_2{}^1||=\left(\!\!\begin{array}{c}3 \\ 7\end{array}\!\!\right)\,,
 ||F_3{}^1||=\left(\!\!\begin{array}{c}3 \\ 7\end{array}\!\!\right)\,,
 ||F_4{}^1||=\left(\!\!\begin{array}{c}1 \\ 2\end{array}\!\!\right)\,.
\end{eqnarray}
Here we can already see that the sum of the degrees of the two hypersurfaces equals the sum of the degree of the third and the fourth $F$. From the last section,  we know how to exchange these functions and how 
to redefine the $\Lambda$'s and $\Gamma$'s in order to obtain 
a sensible new monad. Namely we perform the rescalings
\begin{eqnarray}
\begin{aligned}
 \tilde\Gamma^1 &:=& \left<p_1\right> \Lambda^3,\quad \tilde\Gamma^B &:=& \left<p_1\right> \Lambda^4,\quad
    \tilde \Lambda^3 &:=& \frac{\Gamma^1}{\left<p_1\right>},\quad \tilde \Lambda^4 &:=& \frac{\Gamma^B}{\left<p_1\right>},\\
 \tilde G_1 &:=& F_3{}^1,\quad \tilde G_2 &:=& F_4{}^1,\quad \tilde F_3{}^1 &:=& G_1,\quad \tilde F_4{}^1 &:=& G_2 \,,
\end{aligned}
\end{eqnarray}
yielding the effective superpotential
\begin{eqnarray}\label{eq_dual superpotential example 1} 
\begin{aligned}
\mathcal W = ~\tilde\Gamma^1 \tilde G_1 + \tilde\Gamma^2 \tilde G_2 
+\left<p_1\right>\left( \tilde\Lambda^3 \tilde F_3{}^1 +  \tilde\Lambda^4 \tilde F_4{}^1 + \Lambda^1 F_1{}^1+ \Lambda^2 F_2{}^1\right)\,.
\end{aligned}
\end{eqnarray}
The new charges of the constructed model read
\begin{eqnarray}
\begin{aligned}
 ||\tilde\Gamma^1|| &=&\!\!\!\!   \left(\!\begin{array}{c}-3 \\ -7\end{array}\!\right) ,\quad ||\tilde\Gamma^2|| &=&\!\!\!\! \left(\!\begin{array}{c}-1 \\ -2\end{array}\!\right),\quad
    ||\tilde \Lambda^3|| &=&\!\!\!\! \left(\!\begin{array}{c}1 \\ 4\end{array}\!\right) ,\quad ||\tilde \Lambda^4|| &=&\!\!\!\! \left(\!\begin{array}{c}1 \\ 3\end{array}\!\right),\\
 ||\tilde G_1|| &=&\!\!\!\! \left(\!\begin{array}{c}3
    \\ 7\end{array}\!\right),\quad ||\tilde G_2|| &=&\!\!\!\! \left(\!\begin{array}{c}1 \\ 2\end{array}\!\right),\quad 
 ||\tilde F_3{}^1|| &=&\!\!\!\! \left(\!\begin{array}{c}2 \\ 4\end{array}\!\right),\quad ||\tilde F_4{}^1|| &=&\!\!\!\! \left(\!\begin{array}{c}2 \\ 5\end{array}\!\right) \,
\end{aligned}
\end{eqnarray}
and hence going back to the geometric phase we obtain the new base with a new vector bundle. We notice that the new configuration can be rewritten in a slightly simpler way. The new hypersurface $\tilde G_2$ has precisely the same degree as the divisor $\left\{x_4=0\right\}$ and therefore the corresponding constraining equation to the ambient space simply removes this coordinate from the configuration and we obtain
\begin{equation}
\begin{aligned}
&\begin{array}{|c||c|}
\hline
x_i & \Gamma^j \\
\noalign{\hrule height 1pt}
\begin{array}{ccccccc}
 0 & 0 & 0 & 1 & 1 &  1 \\
 1 & 1 & 1 & 2 & 2 &  0
\end{array}
&
\begin{array}{cc}
 -3   \\
 -7  
 \end{array}\\
\hline
\end{array}
\quad%\\[0.1cm]
&\begin{array}{|c||c|}
\hline
\Lambda^a & p_l \\
\noalign{\hrule height 1pt}
\hline
\begin{array}{cccccccc}
 1 & 0 & 1 & 1  \\
 0 & 1 & 4 & 3 
\end{array}
&
\begin{array}{c}
 -3 \\
 -8  
\end{array}\\
\hline
\end{array}\,.
\end{aligned}
\end{equation}
As was generically shown, this configuration still satisfies the conditions \eqref{eq_anomcancel} and we obtain the following topological data:
\eq{
\label{eq_match chiral spectrum 2}
h_{\widetilde{\mathcal M}}^\bullet(\widetilde{\mathcal V}) &= ( 0, 120 , 0, 0)\,,\\
 h_{\widetilde{\mathcal M}}^{1,1}+h_{\widetilde{\mathcal M}}^{2,1}+h_{\widetilde{\mathcal M}}^1 (\text{End}(\widetilde{\mathcal V})) &=  2 + 95 + 295 = 392\,.
}
If we compare this with the result we obtained in \eqref{eq_match chiral spectrum example 1}, we see that the number of chiral zero modes did not change and the total number of first order deformations stayed the same even though the Hodge number $h^{2,1}$ changed drastically.

Let us once more put some emphasis on the fact that we started up with a base manifold that was of codimension two and due to the exchange ended up with a simpler space given by a codimension one Calabi-Yau manifold. Similarly, as we will see in the next example this can also happen the other way round resulting in an increase of the codimension. Also the number of $C^\ast$ actions can change which will be shown in the following examples, too.\\
\paragraph{\bf An example for an \boldmath{ $SU(4)$}-model:}
Next we present an example of a dual pair of heterotic $(0,2)$ models that give rise to gauge group $SO(10)$ in four dimensions and hence are equipped with a rank 4 vector bundle. The model is again not a deformation of the tangent bundle. The base is the complete intersection of a generic quartic and homogeneous degree hypersurface two inside $\IP^5$. The defining data can be read off in the following table:
\begin{equation}
\begin{aligned}
&\begin{array}{|c||c|}
\hline
x_i & \Gamma^j \\
\noalign{\hrule height 1pt}
\begin{array}{ccccccc}
\IP^5
\end{array}
&
\begin{array}{cc}
 -2 & -4  \\
 \end{array}\\
\hline
\end{array}
%\\[0.1cm]
&\begin{array}{|c||c|}
\hline
\Lambda^a & p_l \\
\noalign{\hrule height 1pt}
\begin{array}{ccccccc}
 1 & 1 & 1 & 1 & 1 & 1 & 1  \\
 \end{array}
&
\begin{array}{ccc}
 -3 & -2 & -2  \\
 \end{array}\\
\hline
\end{array}\,.
\end{aligned}
\end{equation}
\normalsize
\noindent
Clearly this model is anomaly free, i.e.~it satisfies \eqref{eq_anomcancel} and one can also show that the vector bundle is also stable. It is sometimes also referred to as a positive monad, since all line bundles involved have positive degree. It has the following topological data:
\eq{
h_{\widetilde{\mathcal M}}^\bullet(\widetilde{\mathcal V}) &= ( 0, 48 , 0, 0)\,,\\
 h_{\widetilde{\mathcal M}}^{1,1}+h_{\widetilde{\mathcal M}}^{2,1}+h_{\widetilde{\mathcal M}}^1 (\text{End}(\widetilde{\mathcal V})) &=  1 + 89 + 159 = 249\,.
}
Before we move on, we introduce a new coordinate along with a new hypersurface to the model. Doing that at the same time does not change the model at all. In order to perform the exchange of polynomials $F$ and $G$, we have to go to a certain region of the moduli space, exchange them and go back to the generic region in the dual configuration. The resulting base manifold can then be obtained as the conifold transition of the initial base space. The full model is then given by
\begin{equation}
\begin{aligned}
&\begin{array}{|c||c|}
\hline
x_i & \Gamma^j \\
\noalign{\hrule height 1pt}
\begin{array}{cccccccc}
\IP^1\\
\IP^5
\end{array}
&
\begin{array}{ccc}
 -1 & ~0 & -1\\
 -1 & -4 & -1 
 \end{array}\\
\hline
\end{array}
&\begin{array}{|c||c|}
\hline
\Lambda^a & p_l \\
\noalign{\hrule height 1pt}
\begin{array}{ccccccc}
0 & 0 & 0 & 1 & 0 & 0 & 0 \\
1 & 1 & 1 & 0 & 2 & 1 & 1 
 \end{array}
&
\begin{array}{ccc}
 ~0 & -1 & ~ 0  \\
-3 & -2 & -2  
 \end{array}\\
\hline
\end{array}\,,
\end{aligned}
\end{equation}
\normalsize
\noindent
and its topology satisfies the necessary duality check of coinciding spectrum and moduli space dimensions:
\eq{
h_{\widetilde{\mathcal M}}^\bullet(\widetilde{\mathcal V}) &= ( 0, 48 , 0, 0)\,,\\
 h_{\widetilde{\mathcal M}}^{1,1}+h_{\widetilde{\mathcal M}}^{2,1}+h_{\widetilde{\mathcal M}}^1 (\text{End}(\widetilde{\mathcal V})) &=  2 + 86 + 161 = 249\,.
}\\
\paragraph{\bf An example for an \boldmath{ $SU(5)$}-model:}
Finally let us quickly state a different bundle over the same base from the last paragraph. We modify it such that it has no longer $SU(4)$ but rather $SU(5)$ structure. It is given by
\begin{equation}
\begin{aligned}
&\begin{array}{|c||c|}
\hline
x_i & \Gamma^j \\
\noalign{\hrule height 1pt}
\begin{array}{ccccccc}
\IP^5
\end{array}
&
\begin{array}{cc}
 -2 & -4  \\
 \end{array}\\
\hline
\end{array}
&\begin{array}{|c||c|}
\hline
\Lambda^a & p_l \\
\noalign{\hrule height 1pt}
\begin{array}{cccccccc}
 1 & 1 & 1 & 1 & 1 & 1 & 1 & 1\\
 \end{array}
&
\begin{array}{ccc}
 -3 & -3 & -2  
 \end{array}\\
\hline
\end{array}\,.
\end{aligned}
\end{equation}
Since we have still three chiral fields $p_l$ and eight Fermi fields $\Lambda^a$, we end up with a rank 5 vector bundle and hence with an $SU(5)$ gauge group in the four-dimensional  theory. The spectrum and the dimension of the moduli space for this model can be calculated as
\eq{
h_{\widetilde{\mathcal M}}^\bullet(\widetilde{\mathcal V}) &= ( 0, 72 , 0, 0)\,,\\
 h_{\widetilde{\mathcal M}}^{1,1}+h_{\widetilde{\mathcal M}}^{2,1}+h_{\widetilde{\mathcal M}}^1 (\text{End}(\widetilde{\mathcal V})) &=  1 + 89 + 288 = 378\,.
}
The dual base is again given by the same conifold transition as in the last paragraph. Altogether we get
\begin{equation}
\begin{aligned}
&\begin{array}{|c||c|}
\hline
x_i & \Gamma^j \\
\noalign{\hrule height 1pt}
\begin{array}{cccccccc}
\IP^1\\
\IP^5
\end{array}
&
\begin{array}{ccc}
 -1 & ~0 & -1\\
 -1 & -4 & -1 
 \end{array}\\
\hline
\end{array}
%\\[0.1cm]
&\begin{array}{|c||c|}
\hline
\Lambda^a & p_l \\
\noalign{\hrule height 1pt}
\begin{array}{cccccccc}
0 & 0 & 0 & 0 & 0 & 0 & 1 & 0 \\
1 & 1 & 1 & 1  & 1 & 1 & 0 & 2
 \end{array}
&
\begin{array}{ccc}
 ~0 & ~0 &  -1  \\
-3 & -3 & -2  
 \end{array}\\
\hline
\end{array}\,.
\end{aligned}
\end{equation}
\normalsize
\noindent
Calculating the topological data,
\eq{
h_{\widetilde{\mathcal M}}^\bullet(\widetilde{\mathcal V}) &= ( 0, 72 , 0, 0)\,,\\
 h_{\widetilde{\mathcal M}}^{1,1}+h_{\widetilde{\mathcal M}}^{2,1}+h_{\widetilde{\mathcal M}}^1 (\text{End}(\widetilde{\mathcal V})) &=  2 +58 + 318 = 378
}
we can verify that the necessary condition for a duality also holds and hence the conjecture extends to $SU(5)$ bundles as well.
%%%%%%%%%%%%%%%%%%%%%%%%%%%%%%%%%%%%%%%%%%%%%%%
%%%%%%%%%%%%%%%%%%%%%%%%%%%%%%%%%%%%%%%%%%%%%%%
\section{Landscape Scan}
%%%%%%%%%%%%%%%%%%%%%%%%%%%%%%%%%%%%%%%%%%%%%%%
%%%%%%%%%%%%%%%%%%%%%%%%%%%%%%%%%%%%%%%%%%%%%%%
In this section we will review the results that have been obtained in the large landscape scan in \cite{LandscapeStudy}. In contrast to the examples we just discussed, there we focused on models that arise as deformations of the tangent bundle of some Calabi-Yau subvariety in a toric geometry. Here we get the condition \eqref{eq_anomcancel} as well as the bundle stability automatically which saves a lot of work.\\
\paragraph{\bf The scanning algorithm:}
The algorithm that we used to scan through the large sets of configurations was programmed such that it runs through a given set of smooth configurations and produces combinatorially all dual configurations that one obtains by performing the procedure described in \ref{subsec_construction of dual models}. We made sure to start with configurations that are smooth and have a stable bundle. But by constructing the dual models it may happen that new singularities are produced or also that the bundle is destabilized. We did not sufficiently check for these issues but rather made some necessary checks to test it. The algorithm can be summarized by the following chart:
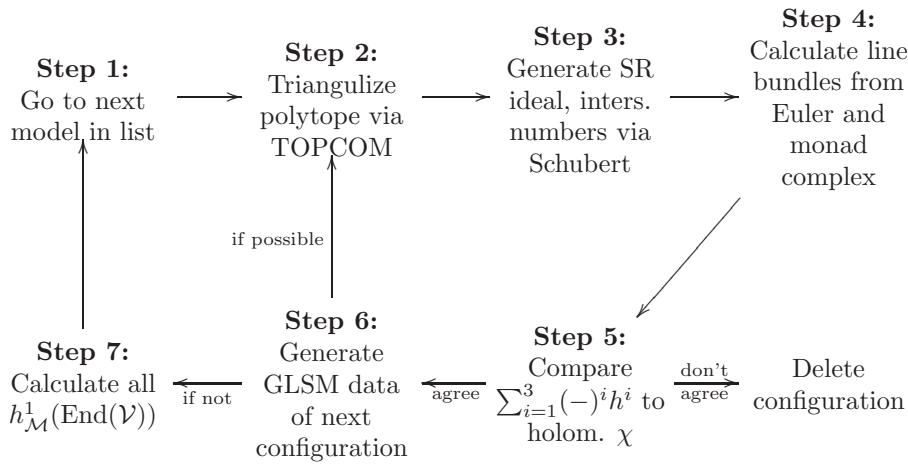
\begin{equation*}
\xymatrix{
  {\parbox{2.3cm}{\center {\bfseries Step 1:}\\ Go to next model in list }} \ar[r]& 
 {\parbox{2.2cm}{\center {\bfseries Step 2:}\\ Triangulize polytope via TOPCOM}}\ar[r] &
 {\parbox{2.2cm}{\center  {\bfseries Step 3:}\\ Generate SR ideal, inters. numbers via Schubert}}\ar[r] & 
 {\parbox{2.2cm}{\center  {\bfseries Step 4:}\\ Calculate line bundles from Euler and monad complex}} \ar[ddl]
 \\ &&& \\
 {\parbox{2.2cm}{\center  {\bfseries Step 7:}\\ Calculate
     all $h^1_{\mathcal M}(\text{End}({\mathcal V}))$}} \ar[uu]& 
 {\parbox{2.2cm}{\center {\bfseries Step 6:}\\ Generate GLSM data of next configuration}} \ar[uu]^{\text{if possible}}\ar[l]^{\text{if not}}& 
 {\parbox{2.3cm}{\center {\bfseries Step 5:}\\ Compare $\sum_{i=1}^3(-)^ih^i$ to holom. $\chi$}} \ar[l]^{\text{agree}}\ar[r]^{\text{don't}}_{\text{agree}}&
 {\parbox{2.2cm}{\center Delete configuration}}
 \\}
\end{equation*}
We ran through two different lists (mentioned in step 1). The first one
contained Calabi-Yau manifolds defined via single hypersurfaces in 
toric varieties. We took
the ambient spaces out of the list from \cite{Kreuzer:2000xy} available on the
website of Maximilian Kreuzer \cite{KreuzerList} and the second list contains
codimension 2 complete intersections in weighed projective spaces which is
part of the list presented in \cite{Klemm:2004km} and available at
\cite{KlemmList}. To resolve the ambient spaces and also to generate the set
of nef partitions to obtain the codimension 2 Calabi-Yaus, we used PALP
\cite{PALP}. For the remaining steps several packages as TOPCOM \cite{TOPCOM}
Schubert \cite{Schubert} and of course
\cohomCalgKoszul~\cite{cohomCalg:Implementation} along with some Mathematica
routines were employed. For the interplay of TOPCOM and Schubert we use the
(not published) Toric Triangulizer \cite{ToricTriangulizer}. \\
\paragraph{\bf The results:}
Our scan ran over the list of hypersurfaces in toric varieties
\cite{KreuzerList} where we considered all toric varieties with 7, 8 and 9
lattice points which make altogether $1,085$. Additionally we scanned over a large set of complete intersections of hypersurfaces in weighted projective spaces. This list can be found online at \cite{KlemmList}. For our scan we simply ran
through the first $2,780$ ambient spaces and chose the $16,029$
possible nef partitions as starting points. All these nef partitions
correspond to  topologically distinct Calabi-Yau manifolds that are complete
intersections of two hypersurfaces in the corresponding weighted projective
space. Starting from a codimension one Calabi-Yau we
performed all first duals to each of those models in the way described in
\ref{subsec_construction of dual models}. Most of the dual models of each hypersurface Calabi-Yau
were given by codimension 2 complete intersections in toric varieties. Since
already many of the duals are obtained by only performing the duality
procedure once, we did not perform duals of duals. Similarly for most of the dual geometries corresponding to a intersection of two hypersurfaces were complete intersections of three hypersurfaces in some toric variety. Some details on the the results can be found in table \ref{tab_results hypersurfaces}. %and a plot of the models that passed the consistency check and arose from codimension one Calabi-Yau spaces can be found in figure \ref{fig_plot hypersurfaces}
\begin{table}%[h]
\begin{center}
\begin{tabular}{c|c|c|c|c|c}
\parbox{0.6cm}{\center Co-dim}&
\parbox{1.5cm}{\center Different classes}&
\parbox{1.5cm}{\center Possibly smooth models} & 
\parbox{1.5cm}{\center Classes without duals}&
\parbox{2cm}{\center Models with matching spectrum}&
\parbox{2cm}{\center Models with full agreement}
%&
%\parbox{2.2cm}{\center Computed (different) line bundle cohom.}
\\
\hline
1&1,085 & 4,507 & 42 & \parbox{2cm}{\center 4,144 (100\%) } & \parbox{2cm}{\center 1509 (94.6\%)} 
%& \parbox{2cm}{\center (1,481,539) 3,069,067}
\\
2&16,961 & 79,204 & 718 & \parbox{2.2cm}{\center 64,332 (85 \%) } & \parbox{2cm}{\center 20,336 (91\%)} 
%& \parbox{2cm}{\center (38,807,002) 109,228,732}
\\
\end{tabular}             \end{center}
\caption{Some data on the landscape study: The codimension is the one of the model we started with. The percent numbers in the parentheses only cover  models where these numbers could actually be calculated. In column 6, by ``full agreement'' we mean that the chiral spectrum of dual models as well as the sum of their complex structure, K\"ahler and bundle deformations agree with the initial ones.}
\label{tab_results hypersurfaces}
\end{table}
%
% \begin{figure}[htbp]
%   \subfigure[Hypersurface models part 1]{
%     %\label{Labelname 1}
% 	 % \includegraphics[angle=0,scale=055]{HyperSurfacePlot1.eps}
% 	      \includegraphics[width=\textwidth]{HyperSurfacePlot1.eps}
%   }
%   \subfigure[Hypersurface models part 2]{
%     %\label{Labelname 1}
% 	 %\includegraphics[angle=0,scale=0.55]{HyperSurfacePlot2.eps}
% 	 \includegraphics[width=\textwidth]{HyperSurfacePlot2.eps}
%   }
% \caption{Plot of the topological data of hypersurfaces in toric varieties and their codimension two duals that passed all necessary consistency checks. Each line corresponds to one class of dual models. Different colored overlapping lines correspond to different classes.}
% \label{fig_plot hypersurfaces}
% \end{figure}
%
%%%%%%%%%%%%%%%%%%%%%%%%%%%%%%%%%%%%%%%%%%%%%%%
%%%%%%%%%%%%%%%%%%%%%%%%%%%%%%%%%%%%%%%%%%%%%%%
\section{Conclusions}
In this letter we reviewed and extended a method to construct from a given heterotic $(0,2)$ model, dual models that generically have the same massless spectra. Continuing earlier work we explicitly showed that this procedure actually also works for models that are not deformations of the tangent bundle and may have $SU(3),~SU(4)$ or $SU(5)$ structure group by testing some necessary conditions for such models to be dual.

We also presented the results of the landscape scan from earlier work which includes many configurations that actually are all deformations of the tangent bundle. They arise as codimension one and codimension two complete intersections in toric varieties and weighted projective spaces respectively. This scan provided evidence for the fact that the proposed procedure generates dual configurations indeed. Having only tested single examples for the scenario where the model is not a deformation of the tangent bundle and therefore comes generically with $SU(n)$ structure for $n=3, 4, 5$ it remains to perform a similar larger scan for such models as well. That we did not do it so far has a reasons. Namely it is first of all not very easy to solve \eqref{eq_anomcancel} in general for a given base geometry. The second thing is that once one has a configuration, one has to check that the bundle is not singular and that it is furthermore stable. Since this is quite a challenge, we did not manage to systematically construct such models, yet. On the other hand, if one could come up with an idea to generate all stable bundles over a given base geometry systematically, it would be no problem to check \eqref{eq_anomcancel} for those models. This was actually already done for a subset of all bundles over specific base spaces \cite{ExploringPositiveMonads} and one way to prove bundle stability in an up to some point systematic way for arbitrary base spaces was suggested in \cite{Anderson:2008ex} and \cite{Anderson:2009sw} and gives hope to enable us to overcome this challenge.
%%%%%%%%%%%%%%%%%%%%%%%%%%%%%%%%%%%%%%%%%%%%%%%
%%%%%%%%%%%%%%%%%%%%%%%%%%%%%%%%%%%%%%%%%%%%%%%
%
%%%%%%%%%%%%%%%%%%%%%%%%%%%%%%%%%%%%%%%%%%%%%%%
%\clearpage
\subsection*{Acknowledgment}
%%%%%%%%%%%%%%%%%%%%%%%%%%%%%%%%%%%%%%%%%%%%%%%
I would like to thank Ralph Blumenhagen and  Benjamin Jurke for discussions and comments as well as Xin Gao for remarks on the article. %I would also like to thank the organizers of the String-Math 2011 conference for putting together such a nice meeting and I am grateful for interesting questions and comments on the results presented there. 
%
%%%%%%%%%%%%%%%%%%%%%%%%%%%%%%%%%%%%%%%%%%%%%%%
\bibliography{rev1}
\bibliographystyle{utphys}
%%%%%%%%%%%%%%%%%%%%%%%%%%%%%%%%%%%%%%%%%%%%%%%                       
\end{document}